\documentclass[graybox]{svmult}

\usepackage{type1cm}        
\usepackage{makeidx}         
\usepackage{graphicx}        
\usepackage{multicol}        
\usepackage[bottom]{footmisc}
\usepackage{hyperref}

\UseRawInputEncoding

\makeindex             

\begin{document}

\title*{Stellar and substellar objects in modified gravity}

\author{Aneta Wojnar}
\institute{Aneta Wojnar \at Institute of Physics, University of Tartu,
W. Ostwaldi 1, 50411 Tartu, Estonia \email{aneta.magdalena.wojnar@ut.ee}
}

\maketitle

\abstract{The last findings on stellar and substellar objects in modified gravity are presented, allowing a reader to quickly jump into this topic. Early stellar evolution of low-mass stars, cooling models of brown dwarfs and giant gaseous exoplanets as well as internal structure of terrestrial planets are discussed. Moreover, possible test of models of gravity with the use of the discussed objects are proposed.}

\section{Basic equations}
\label{sec:1}

There are modifications to the Einstein's gravity which turn out to survive, depending on the features of a given theory of gravity, in the non-relativistic limit derived from their fully relativistic equations.
That is, some of those proposals modify Newtonian gravity, which is commonly used to describe stellar objects, such as the Sun and other stars of the Main Sequence. Those equations are also used to study the substellar family, starting with brown dwarf stars, giant gaseous planets, and even those more similar to the Earth. Therefore, there has appeared a need to explore non-relativistic objects not only for the consistency in describing different astrophysical bodies and gravitational phenomena
with the use of the {\it same} theory of gravity\footnote{however the "which one?" is a question which many physicists try to answer.} but this fact is also an opportunity to understand the nature of the theory, since we better understand the density regimes of such objects. Moreover, since data sets of the discussed stars and exoplanets as well as the accuracy of the observations are still growing, the objects described by non-relativistic equations can be used to constrain some of the gravitational proposals, as presented in the further part of this chapter.

\begin{figure}[t!]
	\centering
	\includegraphics[width=1\linewidth]{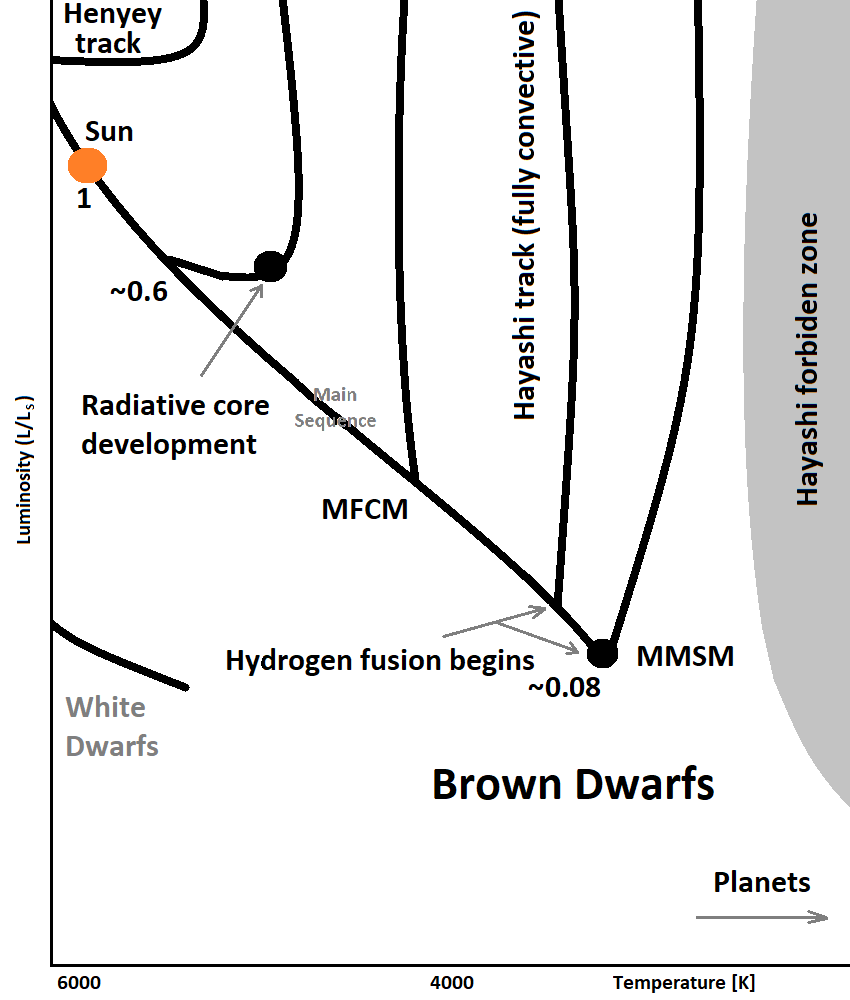}
	\caption{The sketch of the low-temperature region of the evolutionary Hertzsprung-Russell diagram for astrophysical objects discussed in this chapter (the proportions of the evolution and scales are not preserved). A baby star is travelling along the Hayashi track till it reaches the Main Sequence, possibly burning lithium and deuterium. Depending on the star's mass, the object can reach the Main Sequence (MMSM - Minimum Main Sequence Mass indicated as stars with masses $\sim0.08M_\odot$ for hydrogen burning) as a fully convective star (MFCM - Maximal Fully Convective Mass marked), or it can develop a radiative core (it happens for stars with masses $\sim0.6M_\odot$) and then move along the Henyey track. The Hayashi forbidden zone as well as region occupied by brown dwarfs are also indicated. Giant gaseous planets can be found in the colder and dimmer region of the diagram.}
	\label{hr}
\end{figure}

Before discussing the recent findings regarding the topic of non-relativistic objects in modified gravity, we will go through a suitable formalism needed to study low-mass stars and other objects living in the cold and dark edge of the Hertzsprung-Russell diagram (see the picture \ref{hr} and basic literature \cite{stellar,hansen,planets,planets2}). As a working theory we will consider Palatini $f(\bar{R})$ gravity for the Starobinsky model
\begin{equation}
 f(\bar{R})=\bar{R}+\beta\bar{R}^2,
\end{equation}
where $\beta$ is the theory parameter, but similar results as the ones presented here are expected to happen in any theory of gravity which alters Newtonian limit. To read more about Palatini gravity, see \cite{DeFelice:2010aj}, because we will now focus directly on the modified hydrostatic equilibrium equation without its derivation \cite{aneta1,artur,gonzalo,aneta2,aneta3,maria,olek}. Therefore, we will consider a toy-model of a star or planet, that is, a spherical-symmetric low-mass object without taking into account nonsphericity, magnetic fields, and time-dependency, described by the modified hydrostatic equilibrium equation
\begin{equation}\label{pres}
 p'=-g\rho(1+\kappa c^2 \beta [r\rho'-3\rho]) \ ,
\end{equation}
where prime denotes the derivative with respect to the radius coordinate $r$, $\kappa=-8\pi G/c^4$, $G$ and $c$ are Newtonian constant and speed of light, respectively. The quantity $g$ is the surface gravity, approximated on the object's atmosphere as a constant value ($r_{atmosphere}\approx R$, where $R$ is the radius of the object):
\begin{equation}\label{surf}
 g\equiv\frac{G m(r)}{r^2}\sim\frac{GM}{R^2}=\textrm{constant},
\end{equation}
where $M=m(R)$. We will consider only the usual definition for the mass function (however, see the discussion in \cite{olek,olek2} on modified gravity issues)
\begin{equation}\label{masa}
    m'(r)=4\pi r^2\rho(r).
\end{equation}
Using (\ref{masa}) and (\ref{surf}), the equation (\ref{pres}) can be written as
\begin{equation}\label{hyd}
 p'=-g\rho\left( 1+8\beta\frac{g}{c^2 r} \right).
\end{equation}
One of the most important elements in the star's or planet's modelling is the heat transport through object's interior and its atmosphere. A simple and common criterion which determines which kind of the energy transport takes place is given by the Schwarzschild one
 \cite{schw,schw2}:
\begin{eqnarray}
 \nabla_{rad}&\leq&\nabla_{ad}\;\;\textrm{pure diffusive radiative or conductive transport}\\
\nabla_{rad}&>&\nabla_{ad}\;\;\textrm{ adiabatic convection is present locally.}
\end{eqnarray}
The gradient stands for the temperature $T$ variation with depth 
\begin{equation}
 \nabla_{{rad}}:=\left(\frac{d \ln{T}}{d\ln{p}}\right)_{{rad}}.
\end{equation}
The Schwarzschild criterion turns out to be modified in Palatini gravity \cite{aneta2}
\begin{equation}\label{grad}
  \nabla_{rad}=\frac{3\kappa_{rc}lp}{16\pi acG mT^4}\left(1+8\beta\frac{G m}{c^2 r^3}\right)^{-1},
\end{equation}
with $l$ being the local luminosity, the constant  $a=7.57\times 10^{-15}\frac{erg}{cm^3K^4}$ the radiation density while $\kappa_{rc}$ is the radiative and/or conductive opacity. The additional $\beta-$term , depending on the sign of the parameter, has a stabilizing or destabilizing effect. On the other hand, the adiabatic gradient $\nabla_{ad}$ is a constant value for particular cases, as we will see in the further part.

Regarding the microscopic description of matter, an approximation which we will be using here is the polytropic equation of state (EoS): 
\begin{equation}\label{pol}
  p=K\rho^{1+\frac{1}{n}},
\end{equation}
It is good enough for our purposes,
in particularly taking into account the fact that $K$, since it depends on the composition of the fluid, carries information about the interactions between particles, the effects of electron degeneracy, and phase transitions,...\cite{aud}. We will use at least 3 different polytropic EoS, depending on the physical situation. On the other hand, the value of the polytropic index $n$ is related to the class of the astrophysical objects we study \cite{politropia}. The simplest case we will deal with is a fully convective objects with the interior modelled by non-relativistic degenerate
electron gas for which $n=3/2$ while $K$ is given by \cite{stellar}:
\begin{equation}\label{Ka}
    K=\frac{1}{20}\left(\frac{3}{\pi}\right)^\frac{2}{3}\frac{h^2}{m_e}\frac{1}{(\mu_e m_u)^\frac{5}{3}}.
\end{equation}
It is always useful in the case of analytic EoS to write it in the polytropic form (\ref{pol}) since there exists a very convenient approach, called the Lane-Emden (LE) formalism, allowing to rewrite all relevant equations in the dimensionless form. It can be shown that for our particular model of gravity the equation (\ref{hyd}) transforms into the modified Lane-Emden equation \cite{aneta1}
\begin{equation}\label{LE}
 \frac{1}{\xi}\frac{d^2}{d\xi^2}\left[\sqrt{\Phi}\xi\left(\theta-\frac{2\alpha}{n+1}\theta^{n+1}\right)\right]=
 -\frac{(\Phi+\frac{1}{2}\xi\frac{d\Phi}{d\xi})^2}{\sqrt{\Phi}}\theta^n,
\end{equation}
where $\Phi=1+2\alpha \theta^n$ and $\alpha=\kappa c^2\beta\rho_c$. The dimensionless $\theta$ and $\xi$ are defined in the following way 
\begin{eqnarray}
 r=r_c\bar{\xi},\;\;\;\rho=\rho_c\theta^n,\;\;\;p=p_c\theta^{n+1},\;\;\;
 r^2_c=\frac{(n+1)p_c}{4\pi G\rho^2_c},
\end{eqnarray}
with $p_c$ and $\rho_c$ being the core values of pressure and density, respectively. The equation (\ref{LE}) can be solved numerically, and its solution $\theta$ provides star's mass, radius, central density, and temperature: 
\begin{eqnarray}
 M&=&4\pi r_c^3\rho_c\omega_n,\;\;\;
 R=\gamma_n\left(\frac{K}{G}\right)^\frac{n}{3-n}M^\frac{n-1}{n-3},\\
 \rho_c&=&\delta_n\left(\frac{3M}{4\pi R^3}\right) \label{rho0s},\;\;\;
 T=\frac{K\mu}{k_B}\rho_c^\frac{1}{n}\theta_n,
\end{eqnarray}
where $k_B$ is Boltzmann's constant, $\mu$ the mean molecular weight while $\xi_R$ is the dimensionless radius for which $\theta(\xi_R)=0$. In the case of the model of gravity used here the constants (\ref{omega}) and (\ref{delta}) appearing in the above equations also include modifications \cite{artur} but it is not a common feature of modified gravity (see the case of Horndeski gravity, for instance \cite{koyama}, or in Eddington-inspired Born-Infeld gravity, \cite{merce}):
\begin{eqnarray}
 \omega_n&=&-\frac{\xi^2\Phi^\frac{3}{2}}{1+\frac{1}{2}\xi\frac{\Phi_\xi}{\Phi}}\frac{d\theta}{d\xi}\mid_{\xi=\xi_R},\label{omega}\\
  \gamma_n&=&(4\pi)^\frac{1}{n-3}(n+1)^\frac{n}{3-n}\omega_n^\frac{n-1}{3-n}\xi_R,\label{gamma}\\
 \delta_n&=&-\frac{\xi_R}{3\frac{\Phi^{-\frac{1}{2}}}{1+\frac{1}{2}\xi\frac{\Phi_\xi}{\Phi}}\frac{d\theta}{d\xi}\mid_{\xi=\xi_R}}.  \label{delta}
\end{eqnarray}
 Using the LE formalism we may rewrite (\ref{hyd}) and (\ref{grad}) as
 \begin{equation}\label{hyd_pol}
 p'=-g\rho\left( 1-\frac{4\alpha}{3\delta} \right),\;\;\;\;\nabla_{{rad}}=\frac{3\kappa_{rc}lp}{16\pi acG mT^4}\left(1-\frac{4\alpha}{3\delta}\right)^{-1},
\end{equation}
where we have skipped the index $n$ in the parameter $\delta$ (\ref{delta}).
Some of the objects we will consider in those notes are massive enough to burn light elements in their core; it can be either hydrogen, deuterium, or lithium. The product of any of those energy generation processes is luminosity, which can be obtained by the integration of the below expression:
\begin{equation}\label{Lbur}
    \frac{dL_{burning}}{dr}=4\pi r^2\dot\epsilon\rho.
\end{equation}
The energy generation rate $\dot\epsilon$ is a function of energy density, temperature, and stellar composition, however it can be approximated as a power-low function of the two first \cite{fowler}.
The energy produced in the core is radiated through the surface and can be expressed by the Stefan-Boltzmann law 
\begin{equation}\label{stefan}
    L=4\pi f\sigma T_{eff}^4R^2,
\end{equation}
where $\sigma$ is the Stefan-Boltzmann constant. We have added the factor $f\leq1$ which allows to include planets, which obviously radiate less than the black-body with the same effective temperature $T_{eff}$. This particular temperature (as well as other parts of atmosphere modelling) is usually difficult to determine and can carry significant uncertainties. Notwithstanding, there is a tool which we will often use when we look for some characteristics of the atmosphere. It is the optical depth $\tau$, averaged over the object's atmosphere, see e.g.  \cite{stellar,hansen}):
\begin{equation} \label{eq:od}
 \tau(r)=\bar\kappa\int_r^\infty \rho dr,
\end{equation}
where $\bar\kappa$ is a mean opacity. In the further part, since we will mainly work with the objects whose atmospheres have low temperatures, we will use Rosseland mean opacities which are given by the simple Kramers' law
\begin{equation}\label{abs}
 \bar\kappa= \kappa_0 p^u T^{w},
\end{equation}
where $\kappa_0$, $u$ and $w$ are values depending on different opacity regimes \cite{planets,kley}. 
We will also assume that the atmosphere is made of particles satisfying the ideal gas relation ($N_A$ is the Avogardo constant)
\begin{equation}\label{ideal}
 \rho=\frac{\mu p}{N_A k_B T}.
\end{equation}
Again we can use the polytropic EoS (\ref{pol}) to rewrite above as
\begin{equation}\label{eos}
 p=\tilde K T^{1+n},\;\;\;\tilde K=\left(\frac{N_Ak_B}{\mu}\right)^{1+n}K^{-n},
\end{equation}
where $K$ can be shown to be a function of solutions of the modified Lane-Emden equation, an therefore it depends on the theory of gravity \cite{aneta2}.

\section{Pre-Main Sequence phase}
In the following section we will discuss some of the processes related to the early stellar evolution. Before reaching the Main Sequence, a baby star being on the so-called Hayashi track still contracts, decreasing its luminosity but not changing too much its surface temperature. Often the conditions present in the core are sufficient to burn light elements such as deuterium and lithium for instance, however in order to burn hydrogen, the temperature in the star's core must be much higher than in the lithium's case. Moreover, during its journey down along the Hayashi track the pre-Main Sequence star is fully convective apart from its radiative atmosphere. As already mentioned, because of the gravitational contractions the physical conditions in the core are changing and it may happen that the convective core will become radiative. In such a situation, the star will follow subsequently the Henyey track. This phase is much shorten than Hayashi one, it is followed by more massive stars (see the figure \ref{hr}), and will not be discussed here. The radiative core development, hydrogen burning, and other processes related to the early stellar evolution not only depend on the star's mass but also on a theory of gravity, as we will see in the following subsections.

\subsection{Hayashi track}

The photosphere is defined at the radius for which the optical depth (\ref{eq:od}) with mean opacity $\kappa$ is equaled to $2/3$.
Using this relation in order to integrate the hydrostatic equilibrium equation (\ref{hyd}) with $r=R$ and $M=m(R)$, and applying the absorption law (\ref{abs}) for a stellar atmosphere dominated by $H^-$ in the temperature range $3000< T<6000$K with $\kappa_0\approx1.371\times10^{-33}Z\mu^\frac{1}{2}$ and $u=\frac{1}{2},\,w=8.5$, where $Z=0.02$ is solar metallicity \cite{hansen}, one gets
\begin{equation}\label{fotos}
 p_{ph}=8.12\times10^{14}\left(\frac{ M\left(1-\frac{4\alpha}{3\delta}\right)}{L T_{ph}^{4.5}Z\mu^\frac{1}{2}}\right)^\frac{2}{3},
\end{equation}
in which the Stefan-Boltzmann law $L=4\pi\sigma R^2T^4_{ph}$ with $ T_{{eff}{\mid_{r=R}}}\equiv T_{ph}$ was already used.
On the other hand, from (\ref{eos}) taken on the photosphere with $n=3/2$ and applying the Stefan-Boltzmann law again, we have 
\begin{equation}
  T_{ph}=9.196\times10^{-6}\left( \frac{L^\frac{3}{2}Mp_{ph}^2\mu^5}{-\theta'\xi_R^5} \right)^\frac{1}{11}.
\end{equation}
The pressure appearing above is the pressure of the atmosphere; therefore, using (\ref{fotos}) and rescaling mass and luminosity to the solar values $M_\odot$ and 
$L_\odot$, respectively, we can finally write
\begin{equation}\label{hay}
 T_{ph}=2487.77\mu^\frac{13}{51}\left( \frac{L}{L_\odot}  \right)^{\frac{1}{102}}\left( \frac{M}{M_\odot}  \right)^{\frac{7}{51}}
 \left( \frac{\left(\frac{1-\frac{4\alpha}{3\delta}}{Z}\right)^\frac{4}{3}}{\xi_R^5\sqrt{-\theta'}} \right)^\frac{1}{17}\textrm{K}.
\end{equation}
The obtained formula relates the effective temperature and luminosity of the pre-main sequence star for a given mass $M$ and mean molecular weight $\mu$. That is, it provides an evolutionary track called Hayashi track \cite{hayashi}. Those tracks, being almost vertical lines on the right-hand side of the H-R diagram, are followed by the baby stars until they develop the radiative core, or they reach the Main Sequence. Immediately we observe that the effective temperature is nearly constant; but also notice that the temperature coefficient is too low -- it is caused by our toy-model assumptions, mainly related to the atmosphere modelling. However, this simplified analysis allows us to agree that indeed modified gravity shifts the curves (see the figure 2 in \cite{aneta2}), leading to the possibility of constraining models of gravity by studying T Tauri stars \cite{bert} positioned nearby the Hayashi forbidden zone (see the figure \ref{hr}).

\subsection{Lithium burning}
In the fully convective stars (in such a case we may assume that the star is well-mixed) with mass M and hydrogen fraction $X$,
the depletion rate is given by the expression
\begin{equation}\label{reac}
 M\frac{{d}f}{{d}t}=-\frac{Xf}{m_H}\int^M_0\rho\langle\sigma v\rangle dM,
\end{equation}
where the non-resonant reaction rate for the temperature $T<6\times 10^6$ K is
\begin{equation}
 N_A\langle\sigma v\rangle=Sf_{scr} T^{-2/3}_{c6}\textrm{exp}\left[-aT_{c6}^{-\frac{1}{3}}\right]\;\frac{\textrm{cm}^3}{\textrm{s g}},
\end{equation}
where $T_{c6}\equiv T_c/10^6$ K and $f_{scr}$ is the screening correction factor, while $S=7.2\times10^{10}$ and $a=84.72$ are dimensionless parameters 
in the fit to the reaction rate $^7\textrm{Li}(p,\alpha)\,^4\textrm{He}$ \cite{usho,cf,raimann}. The Lane-Emden formalism for Palatini gravity provides the expressions for the central temperature $T_c$ and central density $\rho_c$ (\ref{rho0s}). However, instead of the simplest polytropic model (\ref{Ka}), we need to take into account an arbitrary electron degeneracy degree $\Psi$ and mean molecular weight $\mu_{eff}$, and thus the radius is
\begin{equation}\label{Rpol}
 \frac{R}{R_\odot}\approx\frac{7.1\times10^{-2}\gamma}{\mu_{eff}\mu_e^\frac{2}{3}F^\frac{2}{3}_{1/2}(\Psi)}
 \left(\frac{0.1M_\odot}{M}\right)^\frac{1}{3},
\end{equation}
where $F_n(\Psi)$ is the $n$th order Fermi-Dirac function. Inserting the quantities $T_c$, $\rho_c$, and $R$ given by the Lane-Emden formalism, changing the variables to the spatial ones, and assuming that the burning process is restricted to the central region of the star (so then we can use the near center solution of LE) the depletion rate (\ref{reac}) can be written as \cite{aneta3}
\begin{eqnarray}
 \frac{{d}}{{d}t}\textrm{ln}f&=&-6.54\left(\frac{X}{0.7}\right)\left(\frac{0.6}{\mu_{eff}}\right)^3\left(\frac{0.1M_\odot}{M}\right)^2\nonumber\\
 &\times& Sf_{scr} a^7 u^{-\frac{17}{2}}e^{-u}
 \left(1+\frac{7}{u}\right)^{-\frac{3}{2}}\xi_R^2(-\theta'(\xi_R)),
\end{eqnarray}
where $u\equiv aT_6^{-1/3}$. In order to proceed further, we need to find the dependence on time of the central temperature parameter $u$, which can be obtained from the Stefan-Boltzman equation together with the virial theorem
\begin{equation}
 L=4\pi R^2 T^4_{eff}=-\frac{3}{7}\Omega\frac{GM^2}{R^2}\frac{{d}R}{{d}t}.
\end{equation}
The factor $\Omega$ stands for modified gravity effects on the equation (in Palatini quadratic model $\Omega=1$ for $n=3/2$). The above relations provides the radius and luminosity as functions of time during the contraction phase
\begin{eqnarray}
 \frac{R}{R_\odot}&=&0.85\Omega^\frac{1}{3} \left(\frac{M}{0.1M_\odot}\right)^\frac{2}{3} \left(\frac{3000\textrm{K}}{T_{eff}}\right)^\frac{4}{3}
 \left(\frac{\textrm{Myr}}{t}\right)^\frac{1}{3}\label{Rt}\\
  \frac{L}{L_\odot}&=& 5.25\times10^{-2}\Omega \left(\frac{M}{0.1M_\odot}\right)^\frac{4}{3} \left(\frac{T_{eff}}{3000\textrm{K}}\right)^{\frac{4}{3}}
 \left(\frac{\textrm{Myr}}{t}\right)^\frac{2}{3}\label{Lcontr},
\end{eqnarray}
with the contraction time given as
\begin{eqnarray}\label{tcon}
 t_{cont}\equiv&-&\frac{R}{{d}R/{d}t}\approx841.91  \left(\frac{3000\textrm{K}}{T_{eff}}\right)^4  \left(\frac{0.1M_\odot}{M}\right)\\
 &\times&
  \left(\frac{0.6}{\mu_{eff}}\right)^3 \left(\frac{T_c}{3\times10^6\textrm{K}}\right)^3 \frac{\xi_R^2(-\theta'(\xi_R))\Omega}{\delta^2}\,\textrm{Myr}.\nonumber
\end{eqnarray}
Using equations (\ref{Rt}) and (\ref{Rpol}) it is possible to express the central temperature $T_c$ with the time during the contraction epoch, which results as
\begin{equation}
 \frac{u}{a}=1.15\left(\frac{M}{0.1M_\odot}\right)^{2/9}\left(\frac{\mu_eF_{1/2}(\eta)}{t_6 T^4_{3eff}}\right)^{2/9}
 \times
 \left(\frac{\xi_R^5\Omega^{2/3}(-\theta'(\xi_R))^{2/3}}{\gamma\delta^{2/3}}\right)^{1/3},
\end{equation}
where $T_{3eff}\equiv T_{eff}/3000$K and $t_6\equiv t/10^6$.

Let us focus now on stars with masses $M<0.2M_\odot$ such that the degeneracy effects are insignificant and $\dot{\mu}_{eff}$ can be neglected when compared to $\dot{R}$. Then, we can write the depletion rate as
\begin{eqnarray}
 &\frac{\textrm{dln}f}{\textrm{d} u}& = 1.15\times10^{13}~T_{3eff}^{-4}\left(\frac{X}{0.7}\right)\left(\frac{0.6}{\mu_{eff}}\right)^6
\left(\frac{M_\odot}{M}\right)^3\nonumber\\
&\times& Sf_{scr} a^{16}u^{-\frac{37}{2}}e^{-u}\left(1-\frac{21}{2u}\right)\frac{\xi_R^4(-\theta'(\xi_R))^2\Omega}{\delta^2}.
\end{eqnarray}
The above equation can be integrated from $u_0=\infty$ to $u$ ($\mathcal{F}\equiv\textrm{ln}\frac{f_0}{f}$):
\begin{eqnarray}\label{sol}
 \mathcal{F}=1.15\times10^{13}\frac{X}{0.7}\left(\frac{0.6}{\mu_{eff}}\right)^6
\left(\frac{M_\odot}{M}\right)^3
 \frac{Sf_{scr} a^{16}g(u)}{~T_{3eff}}\frac{\xi_R^4(-\theta'(\xi_R))^2\Omega}{\delta^2},
\end{eqnarray}
where $g(u)=u^{-37/2}e^{-u}-29\Gamma(-37/2,u)$ while $\Gamma(-37/2,u)$ is an upper incomplete gamma function. The $^7\textrm{Li}$ abundance depends on the gravity model.

One obtains the central temperature $T_c$ from $u(\mathcal{F})$ for a given depletion $\mathcal{F}$. The star's age, radius, and luminosity are given by the equations (\ref{tcon}), (\ref{Rt}), and (\ref{Lcontr}). Let us emphasize that all these values depend on the model of gravity, clearly altering the pre-Main Sequence stage of the stellar evolution. Moreover, age determination techniques which are based on lithium abundance measurements are not model-independent: they do depend on a model of gravity used, as presented above (see details in \cite{aneta3}).

\subsection{Approaching the Main Sequence - Hydrogen burning}\label{Shb}
The process of becoming a true star is related to the stable hydrogen burning. It means that the energy produced in this reaction is radiated away through the star's atmosphere, and that the pressure appearing there because of the energy transport balances the gravitational contraction. When a star contracts, the central temperature increases and when it reaches the values $\sim3\times10^6$K in the core, the thermonuclear ignition of hydrogen starts. There are three reactions responsible for this process: $p + p \to d + e^{+}+\nu_e, p +e^{-}+p \to d +\nu_e, p+d \to {^{3}}H_e + \gamma$, where the first one is slow and a bottle-neck for the lower-mass objects; that is, it stands behind the Minimum Main Sequence Mass (MMSM) term. It was demonstrated that the energy generation rate per unit mass for the hydrogen ignition process can be well described by the power law form \cite{fowler,burrows1}
\begin{equation} \label{eq:pp}
\dot{\epsilon}_{pp}= \dot{\epsilon}_c \left(\frac{T}{T_c}\right)^s \left(\frac{\rho}{\rho_c} \right)^{u-1}, \,\,\,\,\,\dot{\epsilon}_c=\epsilon_0T_c^s\rho_c^{u-1} ,
\end{equation}
where
the two exponents can be approximated as  $s \approx 6.31$ and $u \approx 2.28$, while $\dot{\epsilon}_0\approx 3.4\times10^{-9}$ ergs g$^{-1}$s$^{-1}$. For a baby star with the hydrogen fraction $X=0.75$ the number of baryons per electron in low-mass stars is $\mu_e \approx 1.143$.

Using the energy generation rate (\ref{eq:pp}) and luminosity (\ref{Lbur}) formulae, we can integrate the latter over the stellar volume ($M_{-1}=M/(0.1M_\odot)$):
 \begin{equation}\label{lhb}
  \frac{L_{HB}}{L_\odot}=4\pi r_c^3\rho_c\dot{\epsilon}_c\int^{\xi_R}_{0}\xi^2\theta^{n(u+\frac{2}{3}s)}d\xi=
\frac{1.53\times10^7\Psi^{10.15}}{(\Psi+\alpha_d)^{16.46}}
\frac{\delta^{5.487}_{3/2}M^{11.977}_{-1}}{\omega_{3/2}\gamma^{16.46}_{3/2}},
 \end{equation}
where we used the Lane-Emden formalism with
\begin{eqnarray}
 K= \frac{ (3\pi^2)^{2/3} \hbar}{5 m_e m_H^{5/3} \mu_e^{5/3}} \left( 1+ \frac{\alpha_d}{\Psi} \right),
\end{eqnarray}
and the near center solution of the LE equation which is
$\theta(\xi \approx 0)= 1- \frac{\xi^2}{6} \sim \textrm{exp}\left( -\frac{\xi^2}{6}\right)$ for Palatini $f(R)$ gravity. Here, $\alpha_d\equiv5\mu_e/2\mu\approx4.82$.

Now we will focus on finding the photospheric luminosity which must be equaled to (\ref{lhb}) in order to have a star as a stable system.
Therefore, the surface gravity (\ref{surf}) needs to be rewritten wrt the Lane-Emden variables:
\begin{equation}
 g=\frac{3.15\times10^6}{\gamma^2_{3/2}}
M_{-1}^{5/3}\left(1+\frac{\alpha_d}{\Psi}\right)^{-2} \textrm{cm/s}^2.
\end{equation}
The most tricky part is to find the photospheric temperature. Usually it is obtained from matching the specific entropy of the gas and metallic phases of the H-He mixture \cite{burrows1} (here without the phase transition points \cite{aud}) 
\begin{equation}\label{temp}
 T_{ph}=1.8\times10^6\frac{\rho_{ph}^{0.42}}{\Psi^{1.545}}\textrm{K} \ .
\end{equation}
Applying these two results into (\ref{hyd}) and (\ref{ideal}) one writes the photospheric energy density as ($\kappa_{-2}=\kappa_R/(10^{-2}\mathrm{cm^2 g^{-1}})$, $\kappa_R$ is Rosseland's mean opacity):
\begin{equation}
 \frac{\rho_{ph}}{\mathrm{g/cm^3}}=
 5.28\times10^{-5}M^{1.17}_{-1}\left(\frac{1+8\beta\frac{g}{c^2 R} }{\kappa_{-2}}\right)^{0.7}
\frac{\Psi^{1.09}}{\gamma^{1.41}_{3/2}}\left(1+\frac{\alpha_d}{\Psi}\right)^{-1.41}.
\end{equation}
Inserting it into $T_{ph}$ and using the stellar luminosity (\ref{stefan}) we find
\begin{equation}
L_{ph}=28.18L_\odot\frac{M^{1.305}_{-1}}{\gamma^{2.366}_{3/2}\Psi^{4.351}}
 \times\left(\frac{1+8\beta\frac{g}{c^2 R} }{\kappa_{-2}}\right)^{1.183}
 \left(1+\frac{\alpha_d}{\Psi}\right)^{-0.366} \ .
\end{equation}
Finally, writing $L_{HB}=L_{ph}$ and performing non-complicated algebra:
\begin{equation} \label{result}
M_{-1}^{MMSM}=0.290 \frac{\gamma_{3/2}^{1.32} \omega_{3/2}^{0.09}}{\delta_{3/2}^{0.51}} \frac{(\alpha_d + \Psi)^{1.509}}{\Psi^{1.325}} \left(1-1.31\alpha\frac{\left(\frac{\alpha_d+\Psi}{\Psi}\right)^4}{\delta_{3/2}\kappa_{-2} }\right)^{0.111} 
\end{equation}
we have derived the MMSM. It is clearly modified by our model of gravity not only by the parameter $\alpha$, but also the solutions of the LE equation (\ref{LE}).

\section{Low-mass Main Sequence stars}
In every stellar modelling one needs to determine which kind of the energy transport mechanism is present in each particular layer of the given star. It is usually given by the Schwarzschild criterion (\ref{grad}) which is also altered by the model of gravity \cite{aneta2}. Using that result we will demonstrate that the mass limit of fully convective stars on the Main Sequence is shifted and can have a significant effect on how we model the stars from this mass range. Newtonian-based models predict that Main Sequence stars' interiors with masses smaller than $\sim0.6M_\odot$ are fully convective.

Since the star's luminosity decreases when it contracts following the Hayashi track, it may happen that there appears a radiative zone in the star's interior, and then the star will start following the Henyey track \cite{henyey,hen2,hen3}. In the case of the low-mass stars, however, the fully convective baby star may also reach the Main Sequence without developing a radiative core. In order to deal with such a situation, the decreasing luminosity in the Schwarschild condition for the radiative core development condition (it happens when $ \nabla_{rad}=\nabla_{ad}$, where in our simplified model $\nabla_{ad}=0.4$) cannot be lower than the luminosity of H burning (\ref{lhb}). Therefore, the modified Schwarzschild criterion, after inserting (\ref{stefan}) and (\ref{rho0s}) with homology contaction argument, provides the minimum luminosity for the radiative core development:
\begin{equation}\label{lmin}
 L_{min}=9.89\times10^{7}L_\odot  \frac{\delta_{3/2}^{1.064}(\frac{3}{4}\delta_{3/2}-\alpha)}{\xi^{8.67}(-\theta')^{1.73}} \left(\frac{T_{eff}}{\kappa_0}\right)^{0.8} M_{-1}^{4.4},
\end{equation}
where we have used the Kramer's absorption law (\ref{abs}) with $u=1$ and $w=-4.5$. Thus, a star on the onset of the radiative core development will reach the Main Sequence when $L_{min}=L_{HB}$; so the mass of the maximal fully convective star on the Main Sequence is given by the following expression:
\begin{equation}\label{masss}
 M_{-1}=1.7\frac{\mu^{0.9}T_{eff}^{0.11}(\alpha_d+\Psi)^{2.173}}{\Psi^{1.34}\kappa_0^{0.11}}
 \frac{\gamma^{2.173}\omega^{0.132}}{\delta_{3/2}^{0.58}\xi^{1.14}(-\theta')^{0.23}}.
\end{equation}
Let us firstly focus on the GR case, that is, when $\alpha=0$. Considering a star with 
$\alpha_d=4.82$, the degree of the degeneracy electron pressure as $\Psi=9.4$, and the mean molecular weight $\mu=0.618$ with $T_{eff}=4000$K, the maximal mass of the fully convective star on the Main Sequence is:
\begin{equation}
 M=4.86M_\odot\kappa^{-0.11}_0.
\end{equation}
We notice immediately that the final value does depend on the opacity. Considering two Kramers' opacities: the total bound-free and free-free estimated to be (in $\textrm{cm}^2\textrm{g}^{-1}$), \cite{hansen}
\begin{equation}
 \kappa_0^{bf}\approx 4\times10^{25}\mu \frac{ Z(1+X)}{N_Ak_B},\;\;\;\;
  \kappa_0^{ff}\approx 4\times10^{22}\mu\frac{(X+Y)(1+X)}{N_Ak_B},
\end{equation}
the corresponding masses, for $X=0.75$ and $Z=0.02$, are 
\begin{equation}
    M_{bf}=0.099M_\odot,\;\;\;M_{ff}=0.135 M_\odot,
\end{equation}
respectively. The obtained masses, as we expected, are too low - it is a result of our simplified analysis, mainly related to the atmosphere's description and gas behaviour in the considered pressure and temperature regimes. However, we may use the obtained values as reference to compare the result arriving from modified gravity: depending on the parameter's value, the masses can even differ around $50\%$ \cite{aneta2}.

\section{Aborted stars: brown dwarfs}

Let us discuss a family of objects which do not satisfy necessary conditions in their core to ignite hydrogen\footnote{some massive brown dwarfs do burn hydrogen, however the process is not stable ($L_{HB}\neq L_{ph}$) and  since although there is some energy production, the object radiates more than produces, therefore it is cooling down and following the BDs' evolution.} and subsequently to enter the Main Sequence phase. Such an object will radiate away all stored energy, being a result of gravitational contraction and eventual light elements burning in the early stage's evolution. It will stop contracting when the electron degeneracy pressure balances the gravitational pulling, and consequently it will be cooling down with time. In order to study a simple but accurate cooling model of brown dwarfs, we need to consider a more realistic description of matter, as the brown dwarf stars are composed of the mixture of degenerate and ideal gas states at finite temperature. It turns out however that such an EoS can be rewritten in the polytropic form for $n=3/2$ \cite{aud}, but with more complicated polytropic function with $K=C\mu_e^{-\frac{5}{3}}(1+b+a \eta)$, where the constant $C=10^{13} \rm{cm}^4g^{-2/3}s^{-2}$, $a=\frac{5}{2}\mu_e\mu_1^{-1}$, while the number of baryons per electron is represented by $\mu_e$. Here, we use $\eta=\Psi^{-1}$ as the electron degeneracy parameter, while $\mu_1$ takes into account ionization, and it is defined as
\begin{equation}
\frac{1}{\mu_1}=(1+x_{H^+})X+\frac{Y}{4},
\end{equation}
where $x_{H^+}$ is the ionization fraction of hydrogen $X$ ($Y$ stand for helium one) and depends on the phase transitions points \cite{chab4}. Besides, the quantity $b$ is 
\begin{equation}\label{defb}
    b=-\frac{5}{16} \eta \mathrm{ln}(1+e^{-1/\eta})+\frac{15}{8}\eta^2\left( \frac{\pi^2}{3}+\mathrm{Li}_2[-e^{-1  /\eta}] \right),
\end{equation}
where $\rm{Li}_2$ denotes the second order polylogarithm function and the degeneracy parameter is given as $\eta=\frac{k_B T}{\mu_F}$.
Therefore, we can still use the LE formalism for our purposes, that is, we can express the star’s central pressure, radius, central density, and temperature $T_c=\frac{K\mu}{k_B}\rho_c^\frac{1}{n}$ as functions of the above parameters; we will see soon that the degeneracy parameter depends on time because of the still ongoing gravitational contraction.

As already commented, the most uncertain part of our calculations is related to the photospheric values of, for instance, effective temperature. In brown dwarfs' case one usually uses the entropy method, that is, matching the entropy of non-ionized molecular mixture of H and He at the atmosphere to the interior one, composed mainly of degenerate electron gas \cite{aud,burrows1}:
\begin{equation}\label{entr}
 S_{interior}=\frac{3}{2}\frac{k_BN_A}{\mu_{1mod}}(\rm{ln}\eta+12.7065)+C_1,
\end{equation}
where $C_1$ is an integration constant of the first law of thermodynamics and $\mu_{1mod}$ is modified $\mu_1$ at the photosphere (see the details and its form in \cite{maria,aud}. The matching provides the effective temperature as
\begin{equation}\label{tsur}
    T_{eff}=b_1 \times 10^6 \rho_{ph}^{0.4}\eta^\nu\,\,\mathrm{K},
\end{equation}
where the parameters $b_1$ and $\nu$ depend on the specific model describing the phase transition between a metallic H and He state in the BD's interior and the photosphere composed of molecular ones \cite{chab4}. Following the analogous steps as in the section (\ref{Shb}), one gets the photospheric temperature as
\begin{eqnarray}
    T_{eff}=\frac{2.558\times10^4\,\mathrm{K}}{\kappa_R^{0.286}\gamma^{0.572}}
    \left(\frac{M}{M_\odot} \right)^{0.4764}
     \frac{\eta^{0.714\nu}b_1^{0.714}}{(1+b+a\eta)^{0.571}}
    \left(1-1.33\frac{\alpha}{\delta}\right)^{0.286},
\end{eqnarray}
where $\mu_e=1.143$ was used. That allows to find the luminosity of the brown dwarf; hence using the Stefan-Boltzman equation one gets:
\begin{equation}\label{lumph}
    L=\frac{0.0721 L_\odot}{\kappa_R^{1.1424}\gamma^{0.286}}
    \left(\frac{M}{M_\odot} \right)^{1.239}
    \frac{\eta^{2.856\nu}b_1^{2.856}}{(1+b+a\eta)^{0.2848}}
    \left(1-1.33\frac{\alpha}{\delta}\right)^{1.143}.
\end{equation}

The above luminosity depends on time since the electron degeneracy $\eta$ does. To find such a relation for the latter one \cite{burrows1,stev}, let us consider the pace of cooling and contraction given by the first and the second law of thermodynamics
\begin{equation}
    \frac{\rm d E}{\rm d t}+p\frac{\rm d V}{\rm d t}=T\frac{\rm d S}{\rm d t}
    =\dot\epsilon-\frac{\partial L}{\partial M},
\end{equation}
in which the energy generation term $\dot\epsilon$ is negligible in brown dwarfs. We can integrate the above equation over mass to find
\begin{equation}
    \frac{\rm d\sigma}{\rm dt} \left[
    \int N_A k_B T \rm dM
    \right]=-L,
\end{equation}
where $L$ is a surface luminosity and we have defined $\sigma=S/k_BN_A$. The LE polytropic relations allow to get rid of $T$ and $\rho$ and write down
\begin{equation}\label{eqL}
    \frac{\rm d\sigma}{\rm dt}
    \frac{N_A A \mu_e\eta}{C(1+b+a\eta)} \int p \rm dV
   =-L,
\end{equation}
where $A=(3\pi\hbar^3 N_A)^\frac{2}{3}/(2m_e)\approx4.166\times10^{-11}$. The integral in the above equation can be simply found to be $\int p \rm dV=\frac{2}{7}\Omega G\frac{M^2}{R}$ with $\Omega=1$ for $n=3/2$ in Palatini gravity \cite{artur,aneta3}.

With the use of the entropy formula (\ref{entr}) one can easily get the entropy rate as (let us recall that $\sigma=S/k_BN_A$):
\begin{equation}
     \frac{\rm d\sigma}{\rm dt}=\frac{1.5}{\mu_{1{mod}}}\frac{1}{\eta} \frac{\rm d\eta}{\rm dt}.
\end{equation}
Inserting the above expression into (\ref{eqL}) together with the luminosity (\ref{lumph}) gives us the evolutionary equation for the degeneracy parameter $\eta$
\begin{eqnarray}
    \frac{\rm d\eta}{\rm dt}=&-&
    \frac{1.1634\times10^{-18}b_1^{2.856}\mu_{1{mod}}}{\kappa_R^{1.1424}\mu_e^{8/3}}
    \left(\frac{M_\odot}{M} \right)^{1.094}
    \\
    &\times&\eta^{2.856\nu}(1+b+a\eta)^{1.715} \frac{\gamma^{0.7143}}{\Omega} \left(1-1.33\frac{\alpha}{\delta}\right)^{1.143}.\nonumber
\end{eqnarray}
This equation, together with the luminosity equation (\ref{lumph}) and initial conditions $\eta=1$ at $t=0$, provides the cooling process model for a brown dwarf star in Palatini $f(\bar R)$ gravity. To see how modified gravity affects such an evolution after solving these equations numerically\footnote{https://github.com/mariabenitocst/brown$\_$dwarfs$\_$palatini}, see \cite{maria}.

\section{(Exo)-planets}
As we will see, some theories of gravity can change the giant planets' evolution, and may also affect the internal structure of gaseous and terrestrial ones. This fact can change our understanding of the Solar System's formation, as well as it can be used to constrain different gravitational proposals when observational and experimental data with high accuracy are at our disposal. Missions such as ESA's Cosmic Visions \cite{esa} will bring soon more data on the physical properties of Jupiter-like planets, while improved seismic experiments \cite{butler}, as well as those performed in laboratories \cite{merkel}, or with the use of the new generation of the neutrinos’ telescopes \cite{donini} will provide more information about the matter behaviour in the Earth's core and its more exact composition.

\subsection{Jovian planets}

Giant gaseous planets, although their formation processes differs significantly from the one followed by stars and brown dwarfs \cite{planets,planets2}, do also contract and cool down until it reaches the thermal equilibrium, that is, when the received energy from its parent star is equalled to the energy radiated away from the surface of the planet. Their inner description is quite similar to the one of brown dwarfs'; however, the main difference in the cooling process between these two substellar object is that the jovian planets possess an additional source of energy provided by the parent star which cannot be ignored. When a planet with the radius $R_p$ and in the distance $R_{sp}$ from its parent star is in the mentioned thermal equilibrium, it means that its equilibrium temperature 
\begin{equation}
        (1-A_{p})\left(\frac{R_{p}}{2R_{sp}}\right)^2L_{s}=4\pi f\sigma T_{eq}^4R_{p}^2,
\end{equation}
where $A_p$ is an albedo of the planet while $L_s$ the star's luminosity, is equalled to its effective one. However, when we are dealing with some additional energy sources such as for instance  gravitational contraction, Ohmic heating, or tidal forces, it is not so since the planets radiates more than it receives. Therefore, we need a relation between these two temperatures; it is derived from the radiative transport equation with the use of Eddington's approximation \cite{hansen}:
\begin{equation}\label{temp}
 4T^4=3\tau(T^4_{eff}-T^4_{eq})+2(T^4_{eff}+T^4_{eq}),
\end{equation}
where $T$ is the stratification temperature in the atmosphere while $\tau$ is the optical depth. This will allow, when we integrate the equation (\ref{hyd_pol}) with (\ref{abs}), to write down the atmospheric pressure as (see \cite{aneta_jup} for $w=4$): 
\begin{eqnarray}\label{presat}
 p^{u+1}_{w\neq4}=\frac{4^\frac{w}{4}g}{3\kappa_0}\frac{u+1}{1-\frac{w}{4}}\left(1-\frac{4\alpha}{3\delta}\right)
 T_-^{-1}\Big((3\tau T_-+2T_+)^{1-\frac{w}{4}}-(2T_+)^{1-\frac{w}{4}}\Big),
\end{eqnarray}
where we have defined $T_-:=T^4_{eff}-T^4_{eq}$ and $T_+:=T^4_{eff}+T^4_{eq}$. The atmosphere is radiative so there must exist a region in which the convective transport of energy in the planet's interior becomes radiative. In order to find this boundary, we will use the Schwarzschild criterion (\ref{grad}) to find the critical depth in which the radiative process is replaced with the convective one:
\begin{eqnarray}
 \tau_c=\frac{2}{3}\frac{T_+}{T_-}\left(\Big(1+\frac{8}{5}\Big(\frac{\frac{w}{4}-1}{u+1}\Big)\Big)^\frac{1}{\frac{w}{4}-1}-1\right),\;\;w\neq4
\end{eqnarray}
Substituting those expressions into (\ref{presat}) and (\ref{temp}) we may write the formulas for the boundary pressure and temperature
\begin{eqnarray}\label{pbound}
 p^{u+1}_{conv}&=&\frac{8g}{15\kappa_0}\frac{4^\frac{w}{4}\left(1-\frac{4\alpha}{3\delta}\right)}{T_-(2T_+)^{w-1}}\left(\frac{5(u+1)}{5u+8\frac{w}{4}-3}\right),\\
 T^4_{conv}&=&\frac{T_+}{2} \left(\frac{5u+8\frac{w}{4}-3}{5(u+1)}\right)^{\frac{w}{4}-1},\;\;\;\;\;w\neq4.
\end{eqnarray}
On the other hand, to describe the planet's convective interior, let us consider a combination of pressures \cite{don0}
\begin{equation}\label{prescomb}
    p=p_1+p_2,
\end{equation}
where $p_1$ is pressure arising from electron degeneracy, given by the polytropic EoS (\ref{pol}) with $n=3/2$, while $p_2$ is pressure of ideal gas (\ref{ideal}). It can be shown that such a mixture can be again written as a polytrope \cite{stev}. Matching the above interior pressure with (\ref{pbound}) provides a relation between the effective temperature $T_{eff}$ with the radius of the planet $R_p$ which depends on modified gravity:
\begin{eqnarray}\label{cond}
 T_+^{\frac{5}{8}u+\frac{w}{4}-\frac{3}{8}}T_-&=&CG^{-u} M_p^{\frac{1}{3}(2-u)}R_p^{-(u+3)}\mu^{\frac{5}{2}(u+1)}k_B^{-\frac{5}{2}(u+1)}\nonumber\\
 &\times&\gamma^{u+1}(G\gamma^{-1}M_p^\frac{1}{3}R_p-K)^{\frac{5}{2}(u+1)}\left(1-\frac{4\alpha}{3\delta}\right)
\end{eqnarray}
where $C$ is a constant depending on the opacity constants $u$ and $w$:
\begin{equation}
     C_{w\neq4}=\frac{16}{15\kappa_0}2^{\frac{5}{8}(1+u)+\frac{w}{4}}\left(\frac{5u+8\frac{w}{4}-3}{5(u+1)}\right)^{1+\frac{5}{8}(1+u)(\frac{w}{4}-1)}.
\end{equation}
Since the contraction of the planet is a quasi-equilibrium process, the planet's luminosity is a sum of the total energy absorbed by the planet and the internal energy such that for a polytrope with $n=3/2$ \cite{maria} we may write
\begin{equation}\label{cooling}
 L_p=(1-A_{p})\left(\frac{R_{p}}{2R_{sp}}\right)^2L_{s}-\frac{3}{7}\frac{GM_p^2}{R_p^2}\frac{dR_p}{dt}.
\end{equation}
Using (\ref{stefan}), (\ref{abs}) and integrating it from an initial radius $R_0$ to the final one $R_F$, and inserting (\ref{cond}) to get rid of $T_-$ we can derive the cooling equation for jovian planets:
\begin{eqnarray}
  t=-\frac{3}{7}\frac{GM_p^\frac{4}{3}k_B^{\frac{5}{2}(u+1)}\kappa_0}{\pi ac\gamma\mu^{\frac{5}{2}(u+1)}K^{\frac{3}{2}u+\frac{5}{2}}C}\left(1-\frac{4\alpha}{3\delta}\right)^{-1}
 \int^{x_p}_{x_0}\frac{(T_{eff}^4+T^4_{eq})^{\frac{5}{8}u+\frac{w}{4}-\frac{3}{8}}dx}{x^{1-u}(x-1)^{\frac{5}{2}(u+1)}}.\nonumber
\end{eqnarray}
 This, together with (\ref{cond}) providing the effective temperature for a given radius allows to find the age of the planet which clearly differ from the values given by Newtonian physics (see the figure 2 and tables 1-2 in \cite{aneta_jup}).

\subsection{Terrestrial planets}
In this section we will just comment some findings regarding the rocky planets, such as for example the Earth and Mars. Although the numerical analysis demonstrates that we should not expect a large degeneracy in the mass-radius plots for the Earth-sized and smaller planets\footnote{in the case of larger terrestrial planets we observe a significant difference, making the exoplanet's composition more difficult to determine \cite{olek2,seager}.} \cite{olek2} -- however have a look on a more realistic approach in \cite{olek3,olek4} -- it turns out that there is a considerable difference in the density profiles $\rho(r)$, which could be used to constrain and test models of gravity.
Knowing what is the density profile in a given planet allows to obtain the polar moment of inertia $\mathcal{C}$ ($R_p$ is the planet's radius)
\begin{equation}\label{polar}
    \mathcal{C}=\frac{8\pi}{3}\int_0^{R_p} \rho(r) r^4 dr.
\end{equation}
The density profiles provide information on the number of layers composed of different materials (that is, EoS), and their boundaries. The inner structure of the Earth is given by the PREM model \cite{prem,kustowski,iasp91,aki135} being a result of the seismic data analysis, while the martian interior will be known soon, when the Seismic Experiment for Interior Structure from NASA's MARS
InSight Mission's seismometer \cite{nasa} provides the required data.

Since density profiles (central and boundary values of density/pressure, and layers' thickness) are slightly different in modified gravity than those obtained from Newtonian gravity, it means that this fact has an influence on the polar moment of inertia (\ref{polar}), yielding different results for different models of gravity. Such a phenomenon can be compare with the observational value $\mathcal{C}$ provided by precession rate $d\eta/dt$ being caused by gravitational torques from the Sun \cite{kaula}:
\begin{equation}\label{precession}
    \frac{d\eta}{dt}=-\frac{3}{2}J_2\cos{\epsilon}(1-e^2)\frac{n^2}{\omega}\frac{MR^2}{\mathcal{C}}
\end{equation}
where the orbital eccentricity $e$, obliquity $\epsilon$, the rotation rate $\omega$, the effective mean motion $n$ and the gravitational harmonic coefficient $J_2$ are well-known with high accuracy for the Solar System planets, especially for the Earth \cite{ziemia} and Mars \cite{konopliv,smith,folk2}. Therefore, the computed polar moment of inertia from a given model of gravity must agree with the observational one provided by (\ref{precession}). That procedure, when the theoretical modelling improved, can be a powerful tool to test theories of gravity which alters Newtonian equations.

\begin{acknowledgement}
This work was supported by the EU through the European Regional Development Fund CoE program TK133 ``The Dark Side of the Universe".  
\end{acknowledgement}

\end{document}